\documentclass[
]{ceurart}

\usepackage{geometry}
\usepackage{graphicx}
\usepackage{multirow} 
\usepackage{todonotes}
\usepackage{booktabs}
\usepackage{siunitx}
\usepackage[inline]{enumitem}
\usepackage[T1]{fontenc}
\usepackage{lmodern}
\usepackage{etoolbox}
\usepackage{tcolorbox}

\usepackage{listings}
\begin{document}

\copyrightyear{2025}
\copyrightclause{Copyright for this paper by its authors.
Use permitted under Creative Commons License Attribution 4.0
 International (CC BY-SA 4.0).}

\conference{}

\title{Improving LLM Agents with Reinforcement Learning on Cryptographic CTF Challenges}


\author[]{Lajos Muzsai}[
orcid=0009-0007-3009-6225,
email=muzsailajos@protonmail.com,]

\address[]{Eötvös Loránd University, Institute of Mathematics, AI Research Group}

\author[]{David Imolai}[%
orcid=0009-0003-9063-3791,
email=david@imol.ai,
]

\author[]{András Lukács}[%
orcid=0000-0003-3955-9824,
email=andras.lukacs@ttk.elte.hu,
]



\newtcolorbox{promptbox}[1][]{
  colback=gray!5!white,
  colframe=gray!75!black,
  fonttitle=\bfseries,
  title=#1,
  left=2mm, right=2mm, top=1mm, bottom=1mm,
  boxrule=0.5pt, arc=3pt
}



\begin{abstract}
    We present \textsc{Random-Crypto}, a procedurally generated cryptographic Capture The Flag (CTF) dataset designed to unlock the potential of Reinforcement Learning (RL) for LLM-based agents in security-sensitive domains. 
    Cryptographic reasoning offers an ideal RL testbed: it combines precise validation, structured multi-step inference, and reliance on reliable computational tool use. 
    Leveraging these properties, we fine-tune a Python tool-augmented Llama-3.1-8B via Group Relative Policy Optimization (GRPO) in a secure execution environment. 
    The resulting agent achieves a significant improvement in Pass@8 on previously unseen challenges. 
    Moreover, the improvements generalize to two external benchmarks: \emph{picoCTF}, spanning both crypto and non-crypto tasks, and \emph{AICrypto MCQ}, a multiple-choice benchmark of 135 cryptography questions. 
    Ablation studies attribute the gains to enhanced tool usage and procedural reasoning.
    These findings position \textsc{Random-Crypto} as a rich training ground for building intelligent, adaptable LLM agents capable of handling complex cybersecurity tasks.
\end{abstract}



\maketitle

\section{Introduction}

Large Language Models (LLMs) are rapidly evolving from general-purpose text generators into autonomous agents—capable of complex reasoning and increasingly rivaling human performance across domains \cite{llm_agent_survey}. 
This rapid advancement has caused significant research interest in deploying LLM-based agents in critical cybersecurity contexts, such as autonomous penetration testing and vulnerability detection \cite{llm_agents_can_hack_websites, llm_agents_hack_one_day_vuln, team_llm_agent_zero_day}.
Penetration testing, which involves simulating cyberattacks to proactively identify and document exploitable vulnerabilities, is a valuable practice for enhancing system security. Current LLM-based penetration testing agents primarily focus on autonomous exploitation capabilities. Capture The Flag (CTF) challenges have emerged as a natural benchmark for these capabilities: they emulate real-world offensive and defensive scenarios, enforce precise success criteria, and demand structured multi-step reasoning \cite{intercode, hacksynth, nyu, cybench}. Despite advancements in scripted and tool-augmented approaches, a significant challenge remains: the absence of scalable, rigorous environments for training these agents, rather than just evaluating them.

Reinforcement Learning (RL) has emerged as a promising approach to enhance the capabilities of LLM agents and has shown strong results in mathematics, coding, and planning. This enhancement is particularly achieved through structured fine-tuning methods that improve multi-step reasoning and decision-making abilities \cite{llm_rl_survey}. 
The recently proposed Group Relative Policy Optimization (GRPO) \cite{grpo}, in particular, supports efficient fine-tuning using ranked outputs, and has shown strong performance on competitive reasoning tasks with a relatively modest compute effort \cite{deepseek_r1}. However, practical RL training requires environments that provide abundant, objectively verifiable tasks, which are often lacking in cybersecurity datasets. 

We address this gap with \textsc{Random-Crypto}, a procedurally generated dataset for cryptographic Capture The Flag (CTF) challenges that encompasses 50 different algorithmic families and over 5,000 unique tasks. Each challenge is randomized in its parameters, while embedding a randomized text flag, and is set within a contextual narrative generated by an LLM. This benchmark enables agent-centric, reward-driven training at scale, while preserving the symbolic accuracy of cryptographic problem-solving.
This dataset provides an effectively limitless training corpus, allowing for extensive exploration and thorough evaluation of improvements based on RL. By focusing on cryptographic reasoning—a field that involves verifiable objectives, compositional tool use, and step-by-step logical inference—we create an environment where RL can lead agents not only to achieve correctness, but also to develop competence. 

We utilize the \textsc{Random-Crypto} dataset to fine-tune a tool-augmented Llama-3.1-8B model \cite{llama31_8B}, using a secure Python execution environment. 
Our results show that the enhanced agent significantly improves performance on previously unseen crypto tasks, increasing the Pass@8 metric from 0.35 to 0.88. 
Furthermore, the improvements generalize beyond the \textsc{Random-Crypto} distribution to two external benchmarks: \emph{picoCTF}, which spans both cryptographic and non-cryptographic tasks, and \emph{AICrypto MCQ}, a multiple-choice benchmark of 135 cryptography questions covering both theoretical and computational problems. 
In both cases, the model demonstrates reliable, structured tool use learned through reinforcement.

Our key contributions are as follows:
\begin{enumerate}[nosep]
\item We introduce \textsc{Random-Crypto}, a novel procedurally generated cryptographic benchmark for LLM-agent training and evaluation.
\item We demonstrate that GRPO-based RL significantly improves tool-augmented reasoning in LLM agents.
\item We validate cross-domain generalization to \emph{picoCTF} and \emph{AICrypto MCQ}, revealing transferable, RL-acquired strategies for security-relevant problem solving across heterogeneous task formats.
\end{enumerate}

\noindent We have released the benchmarks and training code used for this paper on GitHub.\footnote{https://github.com/aielte-research/HackSynth-GRPO}

\section{Background}

\subsection{CTF Challenges}
CTF challenges are widely used to train and evaluate cybersecurity skills by simulating real-world attack scenarios in sandbox environments. 
The goal is typically to discover a hidden text string, called a flag, embedded in vulnerable applications, binaries, or websites.
CTF challenges encompass a wide array of cybersecurity domains, including: web exploitation, cryptography, reverse engineering, forensics, and binary exploitation.

Cryptographic challenges, in particular, revolve around exploiting weaknesses or logical flaws in encryption and hashing methods. These challenges commonly involve classical ciphers (e.g., Caesar, Vigenère, substitution), symmetric-key encryption (e.g., AES, DES), and public-key encryption schemes (e.g., RSA, ECC). Participants may also encounter tasks involving hashing algorithms, digital signatures, and custom cryptosystems, which often require creative approaches and deep theoretical understanding to solve.

\subsection{LLM Agents in Cybersecurity}

Recent advancements in LLMs have sparked significant interest in their application within cybersecurity, particularly in penetration testing and security assessments.
Researchers have adapted Capture The Flag (CTF) challenges as standardized benchmarks \cite{intercode, hacksynth, cybench, nyu}, to evaluate the cybersecurity capabilities of LLM-based agents systematically. 
Initial evaluations revealed that general-purpose LLMs struggled to effectively handle specialized cybersecurity tasks due to their inherent inability to execute code or interface with operating systems and external tools  \cite{llm_ctf_limitations, intercode, hacksynth, pentestgpt}.

In other domains, to bridge this interaction gap, newer architectures introduced tool augmentation, empowering LLM agents to invoke external tools or APIs. 
Frameworks like ReAct (Reason+Act) demonstrated substantial improvements by integrating logical reasoning with actionable tool use \cite{react}. 
OpenAI’s ChatGPT further extended this approach through a plugin/function-calling API, enabling deterministic JSON tool calls \cite{openai_function_calling}. 
Additionally, Anthropic’s Model Context Protocol (MCP) provides a structured interface for models to interact seamlessly with external data sources and tools, facilitating complex cybersecurity tasks that require real-time computational support \cite{anthropic_mcp, mcp_landscape}.
These tool-augmented LLM agents have significantly advanced the state-of-the-art. However, they have yet to be fully utilized in autonomous penetration testing.

\subsection{Reinforcement Learning to Enhance LLM Reasoning}
Reinforcement Learning (RL) has become a central technique for improving the reasoning abilities of LLMs, particularly in tasks requiring multi-step problem solving and generalization. 
The standard Reinforcement Learning by Human Feedback (RLHF) paradigm fine-tunes models based on human preference signals, typically using Proximal Policy Optimization (PPO) \cite{ppo}, to align model outputs with desired behaviors \cite{train_ppo}.

Recent work has shown that traditional RLHF, while effective in improving helpfulness and safety, often falls short in domains that require structured, multi-hop reasoning like math, programming, and cybersecurity. 
To address this, specialized algorithms have emerged, such as Group Relative Policy Optimization (GRPO) \cite{grpo}. 
GRPO is a variant of PPO tailored for reasoning-intensive tasks: instead of relying on scalar preference labels, it compares multiple candidate outputs and assigns relative rewards within a group.
This relative evaluation strategy encourages diversity and robustness in generated solutions, helping the model refine its internal decision-making process. 
GRPO has been successfully applied to train DeepSeek-R1 \cite{deepseek_r1}, achieving state-of-the-art performance on mathematical reasoning and competitive coding benchmarks. 
Its low computational overhead and compatibility with low-rank adaptation methods such as QLoRA \cite{qlora} also make it an attractive choice for resource-constrained environments.

\section{Methods}

\subsection{\textsc{Random-Crypto} Benchmark Dataset}

We introduce the \textsc{Random-Crypto} benchmark, a programatically generated dataset created for training LLM-based agents with Reinforcement Learning. 
The benchmark was designed to allow generating near infinite variations of CTF challenges based on 50 different cryptographic schemes.
Solving the challenges requires the ability to exploit vulnerabilities in cryptographic schemes and protocols. 
Each challenge requires recovering a hidden flag in the form \texttt{flag\{...\}} by leveraging a specific weakness.

The challenges are categorized into eight archetypes, such as \emph{Classical}, \emph{RSA}, \emph{ECC}, etc., each further divided into multiple sub-types detailed in Appendix Table \ref{app:challenge-taxonomy}. 
These sub-types are partitioned into three difficulty levels:
\begin{itemize}
\item \textbf{Easy}: Simple ciphers solvable manually within minutes without complex scripting (e.g., ROT13 cipher).
\item \textbf{Medium}: Challenges requiring basic scripting and deeper cryptographic understanding (e.g., Morse code decoding with dictionary lookup).
\item \textbf{Hard}: Complex tasks requiring advanced cryptographic knowledge, scripting capabilities, and potentially external online tools or frequency analysis methods (e.g., substitution cipher with frequency analysis or a vulnerable custom cryptosystem).
\end{itemize}
The distribution of subtype difficulties is balanced with at least $16$ challenges from all three difficulty levels in the $50$ subtypes.

Each challenge is generated by first selecting a sub-type and then randomizing parameters specific to the cryptographic scheme, such as cipher keys or substitution dictionaries. 
Subsequently, a random flag is encrypted using the chosen parameters. 
An LLM-generated narrative accompanies each ciphertext, providing context and essential information to ensure solvability (see Appendix~\ref{app:prompts}, Prompt~1). 
For instance, a Caesar cipher challenge randomly selects a shift value between 1 and 25, generates a random plaintext flag, encrypts it, and embeds it within an LLM-generated short narrative. 

To verify the solvability of each challenge, the authors conducted manual validation; at least one instance per sub-type was solved.
Parameter spaces were also reviewed to avoid trivial or unsolvable configurations.
The dataset comprises 50 manually validated challenges used exclusively for testing, along with 5,000 automatically generated challenges designated for agent training.

\subsection{Reinforcement Learning}

We fine-tuned the Llama-3.1-8B-Instruct \cite{llama31_8B} model for 250 training steps using the Group Relative Policy Optimization (GRPO) algorithm. 
The training specifically leveraged a constrained, multi-step code execution interface with a Python execution server via Anthropic's Model Context Protocol (MCP), allowing the model up to four sequential tool invocations per challenge. 
During training, we restricted the training problems to only come from the easy subset of all the challenges.
The decision to restrict training to easy challenges was motivated by initial experiments showing that more difficult challenges negatively impacted the training process.
Models tended to converge prematurely to local optima by prioritizing superficial reward types, i.e., outputs matching formatting or syntax rewards. The rarity of the accuracy reward in the case of these hard challenges caused the models not to pursue solving them. 

During each training step, we generated eight candidate trajectories for four distinct challenges, resulting in a batch size of 32. 
Each candidate trajectory consisted of iterative reasoning followed by structured JSON-formatted tool calls. 
The model interaction cycle included:
\begin{enumerate}[nosep]
\item Generating initial reasoning text,
\item Invoking the Python MCP server with generated code,
\item Receiving execution output and continuing reasoning,
\item Repeating the process until the flag was recovered or the maximum number of interactions (four loops) was reached.
\end{enumerate}

The composite reward function described in Table~\ref{tab:reward-structure} guided the model’s learning process, designed to encourage correct and efficient problem-solving behavior.
To address the challenge of the model prematurely optimizing towards superficial rewards (such as formatting or tool-calling correctness without genuine problem-solving), we implemented a deduction strategy. 
Specifically, we penalized scenarios where the model produced a fictitious or made-up output immediately following a tool call without actually sending it to the MCP server.
Furthermore, we penalized the model when it produced a hallucinated flag as the answer, even before any reasoning or tool calls happened.

The total token generation was limited to 8192 per interaction sequence, and training was augmented using QLoRA \cite{qlora} adapters on one A100 80GB GPU.

We explicitly instructed the model to strictly adhere to the interaction protocol with structured prompts (see Appendix~\ref{app:prompts}, Prompt~2). 
The available Python execution server exposed three functionalities:
\begin{itemize}[nosep]
\item \texttt{execute\_python}: Executes provided Python code and returns output.
\item \texttt{list\_variables}: Lists variables currently stored in memory.
\item \texttt{install\_package}: Installs Python packages into the runtime environment.
\end{itemize}

This structuring mitigated unintended hallucinations by clearly delineating reasoning, tool calls, and output interpretation phases.

\begin{table}[]
  \centering
  \caption{Reward structure used during GRPO training.}
  \label{tab:reward-structure}
  \begin{tabular}{@{}llc@{}}
    \toprule
    \textbf{Reward Type} & \textbf{Description} & \textbf{Value} \\
    \midrule
    Accuracy Reward & The predicted flag fully matches & 1.0 \\
    Answer Format Reward & The final answer matches \texttt{\textbackslash boxed\{flag\{\dots\}\}} & 0.1 \\
    Tool Calling Reward & The model produces a valid JSON tool call & 0.2 \\
    Execution Reward & The MCP server executes the code without error & 0.5 \\ 
    \bottomrule
  \end{tabular}
\end{table}
\hspace{-2em}

\subsection{Evaluation Setup}
We evaluate both baseline and GRPO-trained agents on three benchmarks: 
(i) the \textsc{Random-Crypto} test split, consisting of 50 manually validated cryptographic challenges, 
(ii) the public \emph{picoCTF} benchmark \cite{hacksynth} containing 120 heterogeneous CTF problems across six categories, and 
(iii) the \emph{AICrypto MCQ} benchmark \cite{aicrypto}, a 135-question multiple-choice dataset covering a broad spectrum of cryptography topics, including both theoretical and computational problems.
All evaluations use the same tool-augmented prompting format described in Appendix~\ref{app:prompts}, with eight independent generations per task, and are scored using Pass@8 and Maj@8 metrics.

\section{Results}

This section presents the evaluation of five leading LLMs: Llama-4-Scout-17B-16E \cite{llama4_scout}, Llama-3.1-70B \cite{llama30_70B}, Llama-3.1-8B \cite{llama31_8B}, GPT-4.1 \cite{openai_gpt41}, and o3 \cite{openai_o3}—on the randomized cryptographic benchmark.
Additionally, we analyze the impact of RL-based fine-tuning on model performance.

\subsection{Benchmarking}

To assess the cryptanalytic abilities of contemporary LLMs, we conducted experiments using the 50 verified challenges from the \textsc{Random-Crypto} dataset. 
Each LLM was evaluated under four distinct experimental conditions:
\begin{enumerate*}[label=(\roman*)]
\item \emph{No hint, no tool use} (baseline performance),
\item \emph{Hint only} (supplementary hints provided),
\item \emph{Tool use only} (ability to execute Python code),
\item \emph{Hint and tool use}.
\end{enumerate*}
In each scenario, models received the question field and any supplementary hints or Python tool access. 
Models were required to generate a single, structured response formatted as $\\boxed\{flag{\{\dots}\}\}$ in order for a challenge to be considered solved. 
Due to computational constraints, the maximum token generation was limited to 4096 for the non-tool use case and 8192 for the tool use case.

We report two standard LLM evaluation metrics: Pass@8 is the proportion of tasks for which at least one out of eight independent model generations successfully solves the task, and Maj@8 is the proportion of tasks for which at least five out of eight independent model generations successfully solve the task.
Table \ref{tab:benchmark_results} summarizes the experimental outcomes, measured by Pass@8 and Maj@8 metrics.
GPT-4.1 and o3 models consistently outperform Llama-family models, highlighting the advanced reasoning capabilities of these architectures.

\begin{table}[b]
  \centering
  \caption{Benchmark results (Pass@8/Maj@8) on the \textsc{Random-Crypto} dataset  with and without hint or tool use.}
  \label{tab:benchmark_results}
  \sisetup{table-format=1.2}
  \resizebox{\textwidth}{!}{
  \begin{tabular}{
    l
    *{4}{S}
    *{4}{S}
  }
    \toprule
      & \multicolumn{4}{c}{\textbf{Without Tool Use}}
      & \multicolumn{4}{c}{\textbf{With Tool Use}} \\
    \cmidrule(lr){2-5} \cmidrule(lr){6-9}
      & \multicolumn{2}{c}{\textbf{Without Hint}}
      & \multicolumn{2}{c}{\textbf{With Hint}}
      & \multicolumn{2}{c}{\textbf{Without Hint}}
      & \multicolumn{2}{c}{\textbf{With Hint}} \\
    \cmidrule(lr){2-3}\cmidrule(lr){4-5}
    \cmidrule(lr){6-7}\cmidrule(lr){8-9}
    \textbf{Model} 
      & {Pass@8} & {Maj@8} & {Pass@8} & {Maj@8} 
      & {Pass@8} & {Maj@8} & {Pass@8} & {Maj@8} \\
    \midrule
    Llama-4-Scout-17B-16E
      & 0.08 & 0.02 & 0.14 & 0.02 
      & 0.12 & 0.08 & 0.12 & 0.08 \\
    Llama-3.1-70B       
      & 0.04 & 0.00 & 0.04 & 0.00    
      & 0.18 & 0.02 & 0.12 & 0.00 \\
    Llama-3.1-8B        
      & 0.02 & 0.00 & 0.04 & 0.00    
      & 0.10 & 0.02 & 0.14 & 0.04 \\
    GPT-4.1             
      & 0.26 & 0.14 & 0.26 & \textbf{0.20} 
      & 0.34 & 0.16 & 0.44 & 0.18 \\
    o3              
      & \textbf{0.32} & \textbf{0.20} & \textbf{0.32} & 0.14 
      & \textbf{0.92} & \textbf{0.38} & \textbf{0.92} & \textbf{0.38} \\
    \bottomrule
  \end{tabular}
  }
\end{table}

Interestingly, while supplementary hints typically improve performance for scenarios without tool access, this trend reverses when tools are available, especially for smaller models like Llama-3.1-8B. 
In these cases, hints often prompted models to attempt brute-force approaches. 
For example, challenges requiring enumeration or brute-forcing strategies (such as ROT13 decoding or poorly salted hashes) led smaller models to generate inefficient Python scripts. 
Such attempts often resulted in the models generating Python scripts that tried to enumerate exhaustive lists of possible solutions or unnecessary computations, causing memory exhaustion and crashes in the execution environment.
Conversely, larger and more capable models, such as GPT-4.1 and Scout, demonstrated improved performance when hints were available alongside tools, as they effectively incorporated hints into structured, computationally efficient reasoning processes.

Differences in prompt engineering also affected model performance. 
Specifically, the prompts utilized in this experiment were optimized primarily for Llama-3.1-8B, inadvertently disadvantaging larger Llama-family models.

The o3 model demonstrates exceptional baseline reasoning but experiences notable performance degradation due to the enforced 4096-token limit, cutting short extensive reasoning chains, especially in the presence of hints.
OpenAI's o3 frequently encountered content moderation issues, resulting in lower performance metrics due to incomplete task executions.
The o3 model also consistently got confused regarding permissions for code execution, limiting its effective use of available Python tooling.

\subsection{Reinforcement Learning}

To investigate the impact of Reinforcement Learning (RL) on cryptographic task performance, we conducted three distinct training experiments using the Llama-3.1-8B model:
\begin{enumerate*}[label=(\roman*)]
\item training without hints for 251 steps on easy challenges,
\item training with hints for 251 steps,
\item curriculum-based training, initially providing hints for 130 steps, followed by 121 steps without hints.
\end{enumerate*}

Initially, the model often generated valid Python code through tool calls but would simultaneously provide incorrect or imagined answers instead of executing the code. 
As a result, even when the generated scripts could produce the correct outputs, the model's premature hallucinations led to the termination of the interaction.

\begin{figure}[t]
  \centering
  \resizebox{\textwidth}{!}{
  \includegraphics[width=0.8\linewidth]{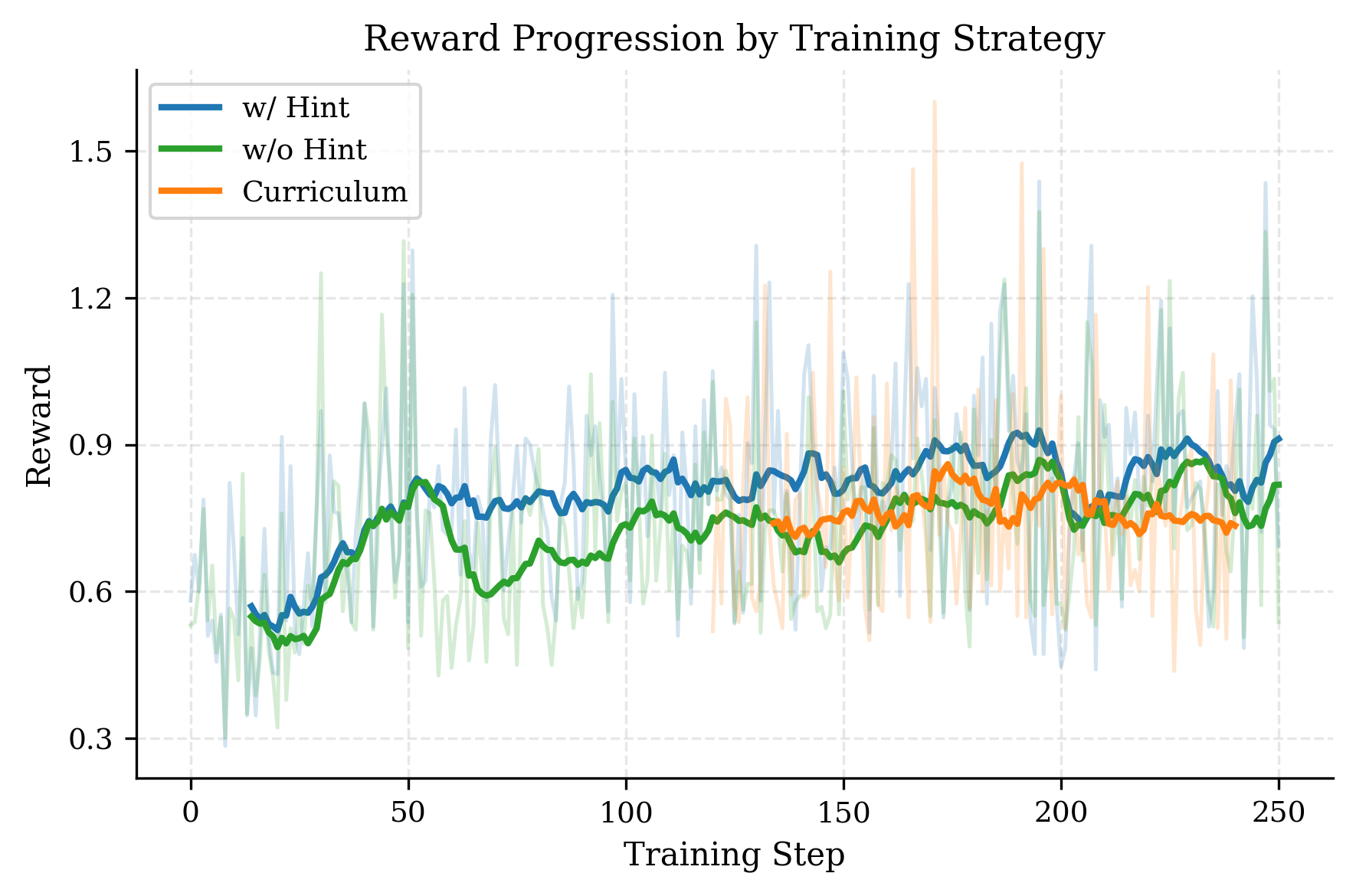}
  }
  \caption{Reward gained during training. The bright lines mark the average, while the shaded lines mark the actual data points. One data point shows the average of all rewards given out in a training step to all benchmarks.}
  \label{fig:reward_plot}
\end{figure}

\begin{figure}[t]
  \centering
  \resizebox{\textwidth}{!}{
  \includegraphics[width=0.8\linewidth]{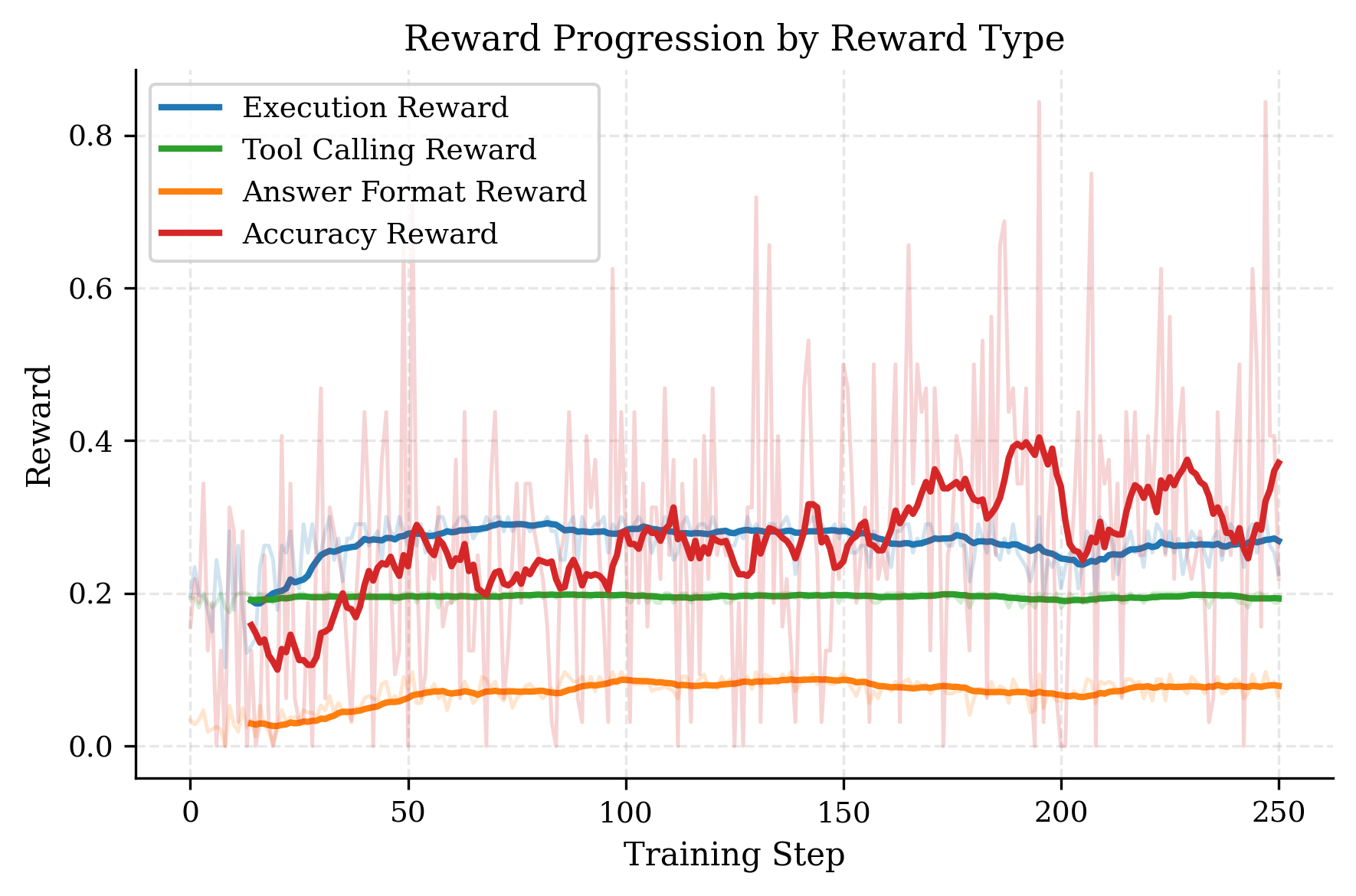}
  }
  \caption{Reward types gained during training. The bright lines mark the average, while the shaded lines mark the actual data points. One data point shows the average of all rewards given out in a training step to all benchmarks. We can observe the biggest threefold improvement in the accuracy of the model, indicating successful challenge resolution.}
  \label{fig:reward_plot_by_types}
\end{figure}

Figure \ref{fig:reward_plot} illustrates the progression of average rewards throughout training, showing an increase from approximately 0.6 to 0.9 across all scenarios. 
The hinted training regime consistently achieved higher rewards due to the hints guiding the model to the appropriate solution.
The exact breakdown of the different rewards gained during training is shown in Figure \ref{fig:reward_plot_by_types}.
Metrics tracking successful Python code execution (defined as code running without errors) improved notably from roughly 39\% to approximately 56\%, with most of this improvement happening in the first 30 steps of training. 
The adherence to the correct answer format saw an even more substantial improvement, rising from about 40\% to nearly 80\%. This improvement happened in the first 50 steps.
However, properly formatted tool calls remained steady at approximately 95\%. The reason behind this is that the prompt was constructed to direct the model to produce proper tool calls.
The rate of successful challenge solutions went from 10\% to 30\%, and unlike the other rewards, this was gradual during the full training, with variance across training steps.

Post-training results on the easy subset of the \textsc{Random-Crypto} benchmark (Table \ref{tab:grpo_crypto_before_after_comparison}) reveal significant performance improvements. 
The curriculum and hinted models both gained a significant 0.80 improvement to their Pass@8 scores, which is better than GPT-4.1 and only slightly behind the o3 model.
The training setup that included no hints achieved a slightly lower Pass@8 score.
On the contrary, it produced a significantly higher Maj@8 score, indicating that this training setup produces a more consistent model.
Training with hints could guide the model toward accurate solutions more frequently, even when the model couldn't guess it by chance. 
However, this could also make the model rely on the guidance too much and lower the performance when no hint is given. 
Overall, results indicate that it is better to leave out hints during the training phase, as the increase in the Pass@8 score is not worth the consistency shown by the Maj@8 score.

\begin{table}[b]
  \centering
  \caption{Pass@8 and Maj@8 results after GRPO training on the \textsc{Random-Crypto}.}
  \label{tab:grpo_crypto_before_after_comparison}
  \sisetup{separate-uncertainty, detect-weight, detect-family}

  \begin{tabular}{
    l
    @{\hspace{12pt}} S[table-format=1.2] S[table-format=1.2]
    @{\hspace{12pt}} S[table-format=+1.2] S[table-format=+1.2]
  }
    \toprule
    & \multicolumn{2}{c}{\textbf{After Training}}
    & \multicolumn{2}{c}{\textbf{Improvement}} \\
    \cmidrule(lr){2-3} \cmidrule(lr){4-5}
    \textbf{Training Type} & \textbf{Pass@8} & \textbf{Maj@8}
                            & \textbf{Pass@8} & \textbf{Maj@8} \\
    \midrule
    w/ Hints    & {\bfseries0.90} & 0.14 & {\bfseries +0.80} & +0.12         \\
    w/o Hints   & 0.88 & {\bfseries0.24} & +0.78 & {\bfseries +0.22}         \\  
    Curriculum  & {\bfseries0.90} & 0.14 & {\bfseries +0.80} & +0.12         \\
    \bottomrule
  \end{tabular}
\end{table}

\subsection{Generalization}
To examine whether the gains obtained on \textsc{Random-Crypto} translate beyond the synthetic distribution, we evaluate the GRPO-fine-tuned agent on the public \emph{picoCTF} benchmark \cite{hacksynth}.
Unlike \textsc{Random-Crypto}, which focuses purely on cryptographic primitives, \emph{picoCTF} spans six heterogeneous categories: \textsc{General}, \textsc{Cryptography}, \textsc{Web}, \textsc{Forensics}, \textsc{Binary Exploitation}, and \textsc{Reverse Engineering}, thereby probing transfer to a substantially broader problem surface. 
This choice is deliberate: success on \emph{picoCTF} requires the agent to (i) recognize when a challenge \emph{can} be reduced to stand-alone Python reasoning, and (ii) invoke the REPL tool robustly under noisier prompts and domain shifts.

Of the 120 \emph{picoCTF} challenges, 95 embed auxiliary artifacts (e.g., ELF binaries, PCAP traces, PNGs) that our current agent cannot yet process efficiently.
Each task is attempted eight times with temperature 0.7 and nucleus sampling ($p=0.95$), and we report Pass@8 and Maj@8 as before.

Table~\ref{tab:grpo_pico} shows that the GRPO model trained with hints or the curriculum training setup increases Pass@8 from 0.10 to 0.18.
Also Maj@8 score was increased by 0.03 with the hinted training setup.
Notably, the training setup without hints achieved the least significant improvement.

\begin{table}[h]
  \centering
  \caption{RL training's effect on performance on the external \emph{picoCTF} benchmark.}
  \label{tab:grpo_pico}
  \sisetup{separate-uncertainty, detect-weight, detect-family}

  \begin{tabular}{
    l
    @{\hspace{12pt}} S[table-format=1.2] S[table-format=1.2]
    @{\hspace{12pt}} S[table-format=+1.2] S[table-format=+1.2]
  }
    \toprule
    & \multicolumn{2}{c}{\textbf{After Training}}
    & \multicolumn{2}{c}{\textbf{Improvement}} \\
    \cmidrule(lr){2-3} \cmidrule(lr){4-5}
    \textbf{Training Type} & \textbf{Pass@8} & \textbf{Maj@8}
                            & \textbf{Pass@8} & \textbf{Maj@8} \\
    \midrule
    w/ Hints     & \textbf{0.18}& \textbf{0.08} & \textbf{+0.11} & \textbf{+0.03} \\
    w/o Hints    & 0.08 & 0.05 & {+0.01} & {+0.00} \\
    Curriculum   & \textbf{0.18} & 0.07 & \textbf{+0.11} & {+0.02} \\
    \bottomrule
  \end{tabular} 
\end{table}


We further evaluated the trained agents on the AICrypto MCQ benchmark \cite{aicrypto}, a 135-question multiple-choice dataset focused exclusively on cryptography. 
Each question offered 4–8 possible answers and spanned a wide range of topics, from purely theoretical items to computational tasks requiring scripting to be solved. 
The model had never been trained on multiple-choice data; however, no filtering or prompt adaptation was applied; we used the same tool-augmented prompt format as in Random-Crypto training. 
As shown in Table \ref{tab:grpo_aicrypto_quiz_comparison}, the curriculum-trained agent achieved the largest improvement over the baseline, increasing Pass@8 by +0.15 and Maj@8 by +0.07, despite the format shift. 
The base model and the other training variants tended to hallucinate the wrong answer after reasoning.
This suggests that curriculum fine-tuning may improve an agent’s adaptability to unseen task formats, balancing both reasoning and computational tool use across heterogeneous cryptographic problems.

\begin{table}[h]
  \centering
  \caption{RL training's effect on performance on the external \emph{AICrypto MCQ} benchmark.}
  \label{tab:grpo_aicrypto_quiz_comparison}
  \sisetup{separate-uncertainty, detect-weight, detect-family}

  \begin{tabular}{
    l
    @{\hspace{12pt}} S[table-format=1.2] S[table-format=1.2]
    @{\hspace{12pt}} S[table-format=+1.2] S[table-format=+1.2]
  }
    \toprule
    & \multicolumn{2}{c}{\textbf{After Training}}
    & \multicolumn{2}{c}{\textbf{Improvement}} \\
    \cmidrule(lr){2-3} \cmidrule(lr){4-5}
    \textbf{Training Type} & \textbf{Pass@8} & \textbf{Maj@8}
                            & \textbf{Pass@8} & \textbf{Maj@8} \\
    \midrule
    w/ Hints     & 0.04 & 0.00 & {+0.00} & {+0.00} \\
    w/o Hints    & 0.03 & 0.00 & {$-0.01$} & {+0.00} \\
    Curriculum   & 0.19 & 0.07 & \textbf{+0.15} & \textbf{+0.07} \\
    \bottomrule
  \end{tabular} 
\end{table}


\section{Discussion}

Executing code generated by LLMs introduces a spectrum of safety and security concerns, particularly in Reinforcement Learning settings where agents are incentivized to solve tasks efficiently. 
In our experiments, an MCP-based Python REPL server \cite{hdresearch_mcp_python} was used to enable persistent tool-use across multiple calls in a single trajectory. 
This design preserved variable state across multiple invocations, simulating a lightweight computation environment with memory.

However, this setup revealed concrete risks. 
We observed that when the agent generated code with excessive memory requirements—such as attempting to enumerate all $6$-character ASCII strings into a list, the REPL server unconditionally executed it. 
As a result, the underlying container crashed due to memory exhaustion. 
This highlights a critical vulnerability in any naive integration of tool-augmented agents with unbounded execution environments.

To mitigate this risk, we strongly advocate for strict containerization and resource sandboxing of all tool-call execution servers. 
This includes enforcing timeouts, memory limits, and instruction whitelisting. 

Beyond resource exhaustion, more subtle alignment failures remain possible. 
While our agents did not attempt to access external URLs, tool APIs with networking capabilities (e.g., web scrapers, HTTP clients) could expose the system to unintentional denial-of-service (DoS) attacks. 
Recent work has begun to explore the risks of LLMs in tool-use scenarios \cite{schick2023toolformer,rai2024guardian,greshake2023not}, but comprehensive sandboxing strategies remain underdeveloped.

\section{Conclusion}

This study demonstrates that Reinforcement Learning can provide Large Language Model (LLM) agents with durable and transferable problem-solving abilities in security-critical domains. 
By leveraging the \textsc{Random-Crypto} benchmark, we fine-tuned a tool-augmented agent using Group Relative Policy Optimization (GRPO) and observed significant improvements in cryptographic reasoning, procedural tool use, and cross-domain generalization.

Our results indicate that RL is not merely optimizing outputs; it enables agents to acquire robust internal routines for reasoning and tool interaction. 
The procedural structure of \textsc{Random-Crypto} played a key role: by offering abundant, verifiable, and tool-centric challenges, it supports scalable, curriculum-aligned agent training.
The gains observed on the AICrypto MCQ benchmark, despite the model never encountering multiple-choice data during training, reinforce that RL-fine-tuning can instill transferable problem-solving routines. 
In particular, the curriculum-trained agent’s advantage here may indicate improved flexibility in adapting tool-augmented reasoning strategies to novel task formats.

Looking ahead, this framework opens the door to agent-centric Reinforcement Learning in cybersecurity and beyond—where competence, adaptability, and safe tool use become foundational design objectives.


\section*{Acknowledgements}

The research was supported by the Hungarian National Research, Development and Innovation Office within the framework of the Thematic Excellence Program 2021 -- National Research Sub programme: ``Artificial intelligence, large networks, data security: mathematical foundation and applications'' and the Artificial Intelligence National Laboratory Program (MILAB).
We would also like to thank GitHub and neptune.ai for providing us with academic access.
Special thanks to Alex Hornyai for helping us validate some of the harder crypto challenges.

\vskip 1.5cm

\bibliography{references}

\begin{thebibliography}{30}
\expandafter\ifx\csname natexlab\endcsname\relax\def\natexlab#1{#1}\fi
\providecommand{\url}[1]{\texttt{#1}}
\providecommand{\href}[2]{#2}
\providecommand{\path}[1]{#1}
\providecommand{\DOIprefix}{doi:}
\providecommand{\ArXivprefix}{arXiv:}
\providecommand{\URLprefix}{URL: }
\providecommand{\Pubmedprefix}{pmid:}
\providecommand{\doi}[1]{\href{http://dx.doi.org/#1}{\path{#1}}}
\providecommand{\Pubmed}[1]{\href{pmid:#1}{\path{#1}}}
\providecommand{\bibinfo}[2]{#2}
\ifx\xfnm\relax \def\xfnm[#1]{\unskip,\space#1}\fi
\bibitem[{Wang et~al.(2024)Wang, Ma, Feng, Zhang, Yang, Zhang, Chen, Tang, Chen, Lin et~al.}]{llm_agent_survey}
\bibinfo{author}{L.~Wang}, \bibinfo{author}{C.~Ma}, \bibinfo{author}{X.~Feng}, \bibinfo{author}{Z.~Zhang}, \bibinfo{author}{H.~Yang}, \bibinfo{author}{J.~Zhang}, \bibinfo{author}{Z.~Chen}, \bibinfo{author}{J.~Tang}, \bibinfo{author}{X.~Chen}, \bibinfo{author}{Y.~Lin}, et~al.,
\newblock \bibinfo{title}{A survey on large language model based autonomous agents},
\newblock \bibinfo{journal}{Frontiers of Computer Science} \bibinfo{volume}{18} (\bibinfo{year}{2024}) \bibinfo{pages}{186345}.
\bibitem[{Fang et~al.(2024{\natexlab{a}})Fang, Bindu, Gupta, Zhan, and Kang}]{llm_agents_can_hack_websites}
\bibinfo{author}{R.~Fang}, \bibinfo{author}{R.~Bindu}, \bibinfo{author}{A.~Gupta}, \bibinfo{author}{Q.~Zhan}, \bibinfo{author}{D.~Kang},
\newblock \bibinfo{title}{{L}{L}{M} agents can autonomously hack websites},
\newblock \bibinfo{journal}{arXiv preprint arXiv:2402.06664}  (\bibinfo{year}{2024}{\natexlab{a}}).
\bibitem[{Fang et~al.(2024{\natexlab{b}})Fang, Bindu, Gupta, and Kang}]{llm_agents_hack_one_day_vuln}
\bibinfo{author}{R.~Fang}, \bibinfo{author}{R.~Bindu}, \bibinfo{author}{A.~Gupta}, \bibinfo{author}{D.~Kang},
\newblock \bibinfo{title}{{L}{L}{M} agents can autonomously exploit one-day vulnerabilities},
\newblock \bibinfo{journal}{arXiv preprint arXiv:2404.08144} \bibinfo{volume}{13} (\bibinfo{year}{2024}{\natexlab{b}}) \bibinfo{pages}{14}.
\bibitem[{Zhu et~al.(2024)Zhu, Kellermann, Gupta, Li, Fang, Bindu, and Kang}]{team_llm_agent_zero_day}
\bibinfo{author}{Y.~Zhu}, \bibinfo{author}{A.~Kellermann}, \bibinfo{author}{A.~Gupta}, \bibinfo{author}{P.~Li}, \bibinfo{author}{R.~Fang}, \bibinfo{author}{R.~Bindu}, \bibinfo{author}{D.~Kang},
\newblock \bibinfo{title}{Teams of {L}{L}{M} agents can exploit zero-day vulnerabilities},
\newblock \bibinfo{journal}{arXiv preprint arXiv:2406.01637}  (\bibinfo{year}{2024}).
\bibitem[{Yang et~al.(2024)Yang, Prabhakar, Narasimhan, and Yao}]{intercode}
\bibinfo{author}{J.~Yang}, \bibinfo{author}{A.~Prabhakar}, \bibinfo{author}{K.~Narasimhan}, \bibinfo{author}{S.~Yao},
\newblock \bibinfo{title}{Intercode: Standardizing and benchmarking interactive coding with execution feedback},
\newblock \bibinfo{journal}{Advances in Neural Information Processing Systems} \bibinfo{volume}{36} (\bibinfo{year}{2024}).
\bibitem[{Muzsai et~al.(2024)Muzsai, Imolai, and Luk{\'a}cs}]{hacksynth}
\bibinfo{author}{L.~Muzsai}, \bibinfo{author}{D.~Imolai}, \bibinfo{author}{A.~Luk{\'a}cs},
\newblock \bibinfo{title}{Hack{S}ynth: {L}{L}{M} agent and evaluation framework for autonomous penetration testing},
\newblock \bibinfo{journal}{arXiv preprint arXiv:2412.01778}  (\bibinfo{year}{2024}).
\bibitem[{Shao et~al.(2024)Shao, Jancheska, Udeshi, Dolan-Gavitt, Xi, Milner, Chen, Yin, Garg, Krishnamurthy et~al.}]{nyu}
\bibinfo{author}{M.~Shao}, \bibinfo{author}{S.~Jancheska}, \bibinfo{author}{M.~Udeshi}, \bibinfo{author}{B.~Dolan-Gavitt}, \bibinfo{author}{H.~Xi}, \bibinfo{author}{K.~Milner}, \bibinfo{author}{B.~Chen}, \bibinfo{author}{M.~Yin}, \bibinfo{author}{S.~Garg}, \bibinfo{author}{P.~Krishnamurthy}, et~al.,
\newblock \bibinfo{title}{{N}{Y}{U} {C}{T}{F} {D}ataset: A scalable open-source benchmark dataset for evaluating {L}{L}{M}s in offensive security},
\newblock \bibinfo{journal}{arXiv preprint arXiv:2406.05590}  (\bibinfo{year}{2024}).
\bibitem[{Zhang et~al.(2024)Zhang, Perry, Dulepet, Jones, Lin, Ji, Menders, Hussein, Liu, Jasper et~al.}]{cybench}
\bibinfo{author}{A.~K. Zhang}, \bibinfo{author}{N.~Perry}, \bibinfo{author}{R.~Dulepet}, \bibinfo{author}{E.~Jones}, \bibinfo{author}{J.~W. Lin}, \bibinfo{author}{J.~Ji}, \bibinfo{author}{C.~Menders}, \bibinfo{author}{G.~Hussein}, \bibinfo{author}{S.~Liu}, \bibinfo{author}{D.~Jasper}, et~al.,
\newblock \bibinfo{title}{Cybench: A framework for evaluating cybersecurity capabilities and risk of language models},
\newblock \bibinfo{journal}{arXiv preprint arXiv:2408.08926}  (\bibinfo{year}{2024}).
\bibitem[{Cao et~al.(2024)Cao, Zhao, Cheng, Shu, Chen, Liu, Liang, Zhao, Yan, and Li}]{llm_rl_survey}
\bibinfo{author}{Y.~Cao}, \bibinfo{author}{H.~Zhao}, \bibinfo{author}{Y.~Cheng}, \bibinfo{author}{T.~Shu}, \bibinfo{author}{Y.~Chen}, \bibinfo{author}{G.~Liu}, \bibinfo{author}{G.~Liang}, \bibinfo{author}{J.~Zhao}, \bibinfo{author}{J.~Yan}, \bibinfo{author}{Y.~Li},
\newblock \bibinfo{title}{Survey on large language model-enhanced reinforcement learning: Concept, taxonomy, and methods},
\newblock \bibinfo{journal}{IEEE Transactions on Neural Networks and Learning Systems}  (\bibinfo{year}{2024}).
\bibitem[{Shao et~al.(2024)Shao, Wang, Zhu, Xu, Song, Bi, Zhang, Zhang, Li, Wu et~al.}]{grpo}
\bibinfo{author}{Z.~Shao}, \bibinfo{author}{P.~Wang}, \bibinfo{author}{Q.~Zhu}, \bibinfo{author}{R.~Xu}, \bibinfo{author}{J.~Song}, \bibinfo{author}{X.~Bi}, \bibinfo{author}{H.~Zhang}, \bibinfo{author}{M.~Zhang}, \bibinfo{author}{Y.~Li}, \bibinfo{author}{Y.~Wu}, et~al.,
\newblock \bibinfo{title}{Deepseekmath: Pushing the limits of mathematical reasoning in open language models},
\newblock \bibinfo{journal}{arXiv preprint arXiv:2402.03300}  (\bibinfo{year}{2024}).
\bibitem[{Guo et~al.(2025)Guo, Yang, Zhang, Song, Zhang, Xu, Zhu, Ma, Wang, Bi et~al.}]{deepseek_r1}
\bibinfo{author}{D.~Guo}, \bibinfo{author}{D.~Yang}, \bibinfo{author}{H.~Zhang}, \bibinfo{author}{J.~Song}, \bibinfo{author}{R.~Zhang}, \bibinfo{author}{R.~Xu}, \bibinfo{author}{Q.~Zhu}, \bibinfo{author}{S.~Ma}, \bibinfo{author}{P.~Wang}, \bibinfo{author}{X.~Bi}, et~al.,
\newblock \bibinfo{title}{Deepseek-r1: Incentivizing reasoning capability in {L}{L}{M}s via reinforcement learning},
\newblock \bibinfo{journal}{arXiv preprint arXiv:2501.12948}  (\bibinfo{year}{2025}).
\bibitem[{{Meta AI}(2024)}]{llama31_8B}
\bibinfo{author}{{Meta AI}}, \bibinfo{title}{{Meta LLaMA 3.1 8B Models}}, \bibinfo{howpublished}{\url{https://huggingface.co/meta-llama/Meta-Llama-3.1-8B-Instruct}}, \bibinfo{year}{2024}. \bibinfo{note}{Accessed: 2025-05-28}.
\bibitem[{Tann et~al.(2023)Tann, Liu, Sim, Seah, and Chang}]{llm_ctf_limitations}
\bibinfo{author}{W.~Tann}, \bibinfo{author}{Y.~Liu}, \bibinfo{author}{J.~H. Sim}, \bibinfo{author}{C.~M. Seah}, \bibinfo{author}{E.-C. Chang},
\newblock \bibinfo{title}{Using large language models for cybersecurity capture-the-flag challenges and certification questions},
\newblock \bibinfo{journal}{arXiv preprint arXiv:2308.10443}  (\bibinfo{year}{2023}).
\bibitem[{Deng et~al.(2023)Deng, Liu, Mayoral-Vilches, Liu, Li, Xu, Zhang, Liu, Pinzger, and Rass}]{pentestgpt}
\bibinfo{author}{G.~Deng}, \bibinfo{author}{Y.~Liu}, \bibinfo{author}{V.~Mayoral-Vilches}, \bibinfo{author}{P.~Liu}, \bibinfo{author}{Y.~Li}, \bibinfo{author}{Y.~Xu}, \bibinfo{author}{T.~Zhang}, \bibinfo{author}{Y.~Liu}, \bibinfo{author}{M.~Pinzger}, \bibinfo{author}{S.~Rass},
\newblock \bibinfo{title}{Pentest{G}{P}{T}: An {L}{L}{M}-empowered automatic penetration testing tool},
\newblock \bibinfo{journal}{arXiv preprint arXiv:2308.06782}  (\bibinfo{year}{2023}).
\bibitem[{Yao et~al.(2023)Yao, Zhao, Yu, Du, Shafran, Narasimhan, and Cao}]{react}
\bibinfo{author}{S.~Yao}, \bibinfo{author}{J.~Zhao}, \bibinfo{author}{D.~Yu}, \bibinfo{author}{N.~Du}, \bibinfo{author}{I.~Shafran}, \bibinfo{author}{K.~Narasimhan}, \bibinfo{author}{Y.~Cao},
\newblock \bibinfo{title}{React: Synergizing reasoning and acting in language models},
\newblock in: \bibinfo{booktitle}{International Conference on Learning Representations (ICLR)}, \bibinfo{year}{2023}.
\bibitem[{{OpenAI}(2024)}]{openai_function_calling}
\bibinfo{author}{{OpenAI}}, \bibinfo{title}{Function calling | {O}pen{A}{I} {P}latform}, \bibinfo{year}{2024}. \URLprefix \url{https://platform.openai.com/docs/guides/function-calling}, \bibinfo{note}{accessed: 2025-05-28}.
\bibitem[{{Anthropic}(2024)}]{anthropic_mcp}
\bibinfo{author}{{Anthropic}}, \bibinfo{title}{Model {C}ontext {P}rotocol: An open standard for connecting {A}{I} models to tools and data}, \bibinfo{year}{2024}. \URLprefix \url{https://www.anthropic.com/news/model-context-protocol}, \bibinfo{note}{accessed: 2025-05-28}.
\bibitem[{Hou et~al.(2025)Hou, Zhao, Wang, and Wang}]{mcp_landscape}
\bibinfo{author}{X.~Hou}, \bibinfo{author}{Y.~Zhao}, \bibinfo{author}{S.~Wang}, \bibinfo{author}{H.~Wang},
\newblock \bibinfo{title}{Model context protocol ({MCP}): Landscape, security threats, and future research directions},
\newblock \bibinfo{journal}{arXiv preprint arXiv:2503.23278}  (\bibinfo{year}{2025}).
\bibitem[{Schulman et~al.(2017)Schulman, Wolski, Dhariwal, Radford, and Klimov}]{ppo}
\bibinfo{author}{J.~Schulman}, \bibinfo{author}{F.~Wolski}, \bibinfo{author}{P.~Dhariwal}, \bibinfo{author}{A.~Radford}, \bibinfo{author}{O.~Klimov},
\newblock \bibinfo{title}{Proximal policy optimization algorithms},
\newblock \bibinfo{journal}{arXiv preprint arXiv:1707.06347}  (\bibinfo{year}{2017}).
\bibitem[{Ouyang et~al.(2022)Ouyang, Wu, Jiang, Almeida, Wainwright, Mishkin, Zhang, Agarwal, Slama, Ray et~al.}]{train_ppo}
\bibinfo{author}{L.~Ouyang}, \bibinfo{author}{J.~Wu}, \bibinfo{author}{X.~Jiang}, \bibinfo{author}{D.~Almeida}, \bibinfo{author}{C.~Wainwright}, \bibinfo{author}{P.~Mishkin}, \bibinfo{author}{C.~Zhang}, \bibinfo{author}{S.~Agarwal}, \bibinfo{author}{K.~Slama}, \bibinfo{author}{A.~Ray}, et~al.,
\newblock \bibinfo{title}{Training language models to follow instructions with human feedback},
\newblock \bibinfo{journal}{Advances in Neural Information Processing Systems} \bibinfo{volume}{35} (\bibinfo{year}{2022}) \bibinfo{pages}{27730--27744}.
\bibitem[{Dettmers et~al.(2023)Dettmers, Pagnoni, Holtzman, and Zettlemoyer}]{qlora}
\bibinfo{author}{T.~Dettmers}, \bibinfo{author}{A.~Pagnoni}, \bibinfo{author}{A.~Holtzman}, \bibinfo{author}{L.~Zettlemoyer},
\newblock \bibinfo{title}{Qlora: Efficient finetuning of quantized {L}{L}{M}s},
\newblock \bibinfo{journal}{Advances in Neural Information Processing Systems} \bibinfo{volume}{36} (\bibinfo{year}{2023}) \bibinfo{pages}{10088--10115}.
\bibitem[{Wang et~al.(2025)Wang, Liu, Ji, Luo, Li, Zhou, Feng, Wang, Cao, Zhang et~al.}]{aicrypto}
\bibinfo{author}{Y.~Wang}, \bibinfo{author}{Y.~Liu}, \bibinfo{author}{L.~Ji}, \bibinfo{author}{H.~Luo}, \bibinfo{author}{W.~Li}, \bibinfo{author}{X.~Zhou}, \bibinfo{author}{C.~Feng}, \bibinfo{author}{P.~Wang}, \bibinfo{author}{Y.~Cao}, \bibinfo{author}{G.~Zhang}, et~al.,
\newblock \bibinfo{title}{Aicrypto: A comprehensive benchmark for evaluating cryptography capabilities of large language models},
\newblock \bibinfo{journal}{arXiv preprint arXiv:2507.09580}  (\bibinfo{year}{2025}).
\bibitem[{{Meta AI}(2025)}]{llama4_scout}
\bibinfo{author}{{Meta AI}}, \bibinfo{title}{{Meta LLaMA-4-Scout-17B-16E Model}}, \bibinfo{howpublished}{\url{https://huggingface.co/meta-llama/Llama-4-Scout-17B-16E-Instruct}}, \bibinfo{year}{2025}. \bibinfo{note}{Accessed: 2025-05-28}.
\bibitem[{{Meta AI}(2024)}]{llama30_70B}
\bibinfo{author}{{Meta AI}}, \bibinfo{title}{{Meta LLaMA 3.1 70B Model}}, \bibinfo{howpublished}{\url{https://huggingface.co/meta-llama/Meta-Llama-3.1-70B-Instruct}}, \bibinfo{year}{2024}. \bibinfo{note}{Accessed: 2025-05-28}.
\bibitem[{OpenAI(2025{\natexlab{a}})}]{openai_gpt41}
\bibinfo{author}{OpenAI}, \bibinfo{title}{{GPT-4.1 Model}}, \bibinfo{howpublished}{\url{https://openai.com/index/gpt-4-1/}}, \bibinfo{year}{2025}{\natexlab{a}}. \bibinfo{note}{Accessed: 2025-05-28}.
\bibitem[{OpenAI(2025{\natexlab{b}})}]{openai_o3}
\bibinfo{author}{OpenAI}, \bibinfo{title}{Openai o3 model}, \bibinfo{howpublished}{\url{https://openai.com/index/introducing-o3-and-o4-mini/}}, \bibinfo{year}{2025}{\natexlab{b}}. \bibinfo{note}{Accessed: 2025-05-28}.
\bibitem[{Research(2025)}]{hdresearch_mcp_python}
\bibinfo{author}{H.~D. Research}, \bibinfo{title}{{MCP-Python}: A {P}ython {REPL} server for the {M}odel {C}ontext {P}rotocol}, \bibinfo{howpublished}{\url{https://github.com/hdresearch/mcp-python}}, \bibinfo{year}{2025}. \bibinfo{note}{Accessed: 2025-05-29}.
\bibitem[{Schick et~al.(2023)Schick, Dwivedi-Yu, Dess{\`\i}, Raileanu, Lomeli, Hambro, Zettlemoyer, Cancedda, and Scialom}]{schick2023toolformer}
\bibinfo{author}{T.~Schick}, \bibinfo{author}{J.~Dwivedi-Yu}, \bibinfo{author}{R.~Dess{\`\i}}, \bibinfo{author}{R.~Raileanu}, \bibinfo{author}{M.~Lomeli}, \bibinfo{author}{E.~Hambro}, \bibinfo{author}{L.~Zettlemoyer}, \bibinfo{author}{N.~Cancedda}, \bibinfo{author}{T.~Scialom},
\newblock \bibinfo{title}{Toolformer: Language models can teach themselves to use tools},
\newblock \bibinfo{journal}{Advances in Neural Information Processing Systems} \bibinfo{volume}{36} (\bibinfo{year}{2023}) \bibinfo{pages}{68539--68551}.
\bibitem[{Rai et~al.(2024)Rai, Sood, Madisetti, and Bahga}]{rai2024guardian}
\bibinfo{author}{P.~Rai}, \bibinfo{author}{S.~Sood}, \bibinfo{author}{V.~K. Madisetti}, \bibinfo{author}{A.~Bahga},
\newblock \bibinfo{title}{Guardian: A multi-tiered defense architecture for thwarting prompt injection attacks on {LLM}s},
\newblock \bibinfo{journal}{Journal of Software Engineering and Applications} \bibinfo{volume}{17} (\bibinfo{year}{2024}) \bibinfo{pages}{43--68}.
\bibitem[{Greshake et~al.(2023)Greshake, Abdelnabi, Mishra, Endres, Holz, and Fritz}]{greshake2023not}
\bibinfo{author}{K.~Greshake}, \bibinfo{author}{S.~Abdelnabi}, \bibinfo{author}{S.~Mishra}, \bibinfo{author}{C.~Endres}, \bibinfo{author}{T.~Holz}, \bibinfo{author}{M.~Fritz},
\newblock \bibinfo{title}{Not what you've signed up for: Compromising real-world {LLM}-integrated applications with indirect prompt injection},
\newblock in: \bibinfo{booktitle}{Proceedings of the 16th ACM Workshop on Artificial Intelligence and Security}, \bibinfo{year}{2023}, pp. \bibinfo{pages}{79--90}.

\end{thebibliography}

\newpage
\appendix
\section{Additional Dataset Details}

Table~\ref{app:challenge-taxonomy} provides a full breakdown of the challenge taxonomy used in our benchmark.
We organized the cryptographic CTF challenges into eight high-level archetypes based on the underlying encryption techniques. 
Each archetype contains several subtypes, which represent specific cryptographic mechanisms or exploit strategies.

\begin{table*}[h]
\centering
\caption{Taxonomy of Procedural CTF Challenge Types}
\label{app:challenge-taxonomy}
\begin{tabular}{p{0.2\linewidth} | p{0.72\linewidth}}
\toprule
\textbf{Archetype} & \textbf{Subtypes} \\
\midrule
\textbf{Classical} & 
Caesar, Vigenère, Playfair, Hill, Rail fence, Substitution, Substitution\_direct, Transposition, Autokey, Atbash, XOR, Hex, ASCII shift, Morse code, Fibonacci encoding, Base64, Base64\_layered, Base85, Base85\_layered, Split flag, Reversed flag, Chunked flag \\
\midrule
\textbf{RSA} & 
Small primes, Repeated prime usage, Partial key exposure, Common factors, Shared prime, Blum integers \\
\midrule
\textbf{AES} & 
AES-GCM, AES-CCM, AES-XTS, AES-CFB \\
\midrule
\textbf{ECC} & 
Small-order curves, Faulty curve parameters, Reused nonce (ECDSA) \\
\midrule
\textbf{Hash} & 
MD5 reverse, Poor random salt, Iterated hash challenge \\
\midrule
\textbf{PRNG} & 
Predictable seed, Time-based seed, Low-entropy generator, LFSR weakness, Congruential generator flaw \\
\midrule
\textbf{Web Crypto} & 
JWT ‘none' algorithm, Weak cookie encryption, Broken key exchange, Insecure session token \\
\midrule
\textbf{Signature Schemes} & 
ECDSA nonce reuse, RSA sign with low public exponent \\
\bottomrule
\end{tabular}
\end{table*}

\section{Prompts used in the paper}
\label{app:prompts}

\subsection*{Prompt 1: Challenge Story Generation Prompt}
The following prompt was used to instruct the LLM to generate a short narrative for each cryptographic challenge. 
The variable \texttt{CIPHER} is replaced with the name of the cipher used in the instance, and \texttt{VULNERABILITY} is an optional string describing the nature of the weakness being exploited.

\begin{promptbox}[Prompt 1: Challenge Story Generation Prompt]
Write a \textbf{very short} ($\leq$ 30 words) back-story for a crypto
challenge that uses the \texttt{CIPHER} cipher \texttt{VULNERABILITY}.

Keep it \textbf{very brief} without unnecessary information.

Return \textbf{only} the text between these XML tags:
<challenge>…</challenge>

Insert the placeholder <CIPHER> \textbf{verbatim} where the ciphertext will go.
Do \textbf{not} write anything outside the tags.
\end{promptbox}

\subsection*{Prompt 2: Tool-Calling Agent Prompt}

This prompt instructs the LLM to perform tool-augmented reasoning using the \texttt{execute\_python} tool. 
The placeholder \texttt{MCP\_TOOL\_LIST} is replaced at runtime with a list of available tools, and \texttt{QUESTION} is the current question to solve.

\begin{promptbox}[Prompt 2: Tool-Augmented Agent Instructions]
You are an assistant with access to the tools listed below.

\vspace{1em}
\textbf{Rules (follow all of them):}

1. Do your chain-of-thought inside \texttt{<reasoning> ... </reasoning>} tags.

2. \textbf{If} you need to run Python, output \textbf{EXACTLY ONE} JSON object \textbf{and nothing else} in that assistant message.  
   
   The object \emph{must} have the shape

       \{
       
        \hspace*{2em}"name": "execute\_python",
         
         \hspace*{2em}"inputs": \{
         
           \hspace*{4em}"code": "print('hello')",
           
           \hspace*{4em}"reset": false
           
         \hspace*{2em}\}
         
       \}

   – no prose before or after, no Markdown formatting.
   
3. After the tool reply arrives (role \texttt{"tool"}), read its \texttt{Output:} and then answer the
   question. 
   
   Wrap the final flag in \texttt{\textbackslash boxed\{flag\{...\}\}}.  
   
   \textbf{Never} guess the flag before you have seen it in a tool response.
   
4. \textbf{Never} include \texttt{\textbackslash boxed\{flag\{...\}\}} in the SAME assistant message that
   contains a JSON tool call. 
   
   Only write the boxed flag \textbf{after} the
   tool’s reply (\texttt{role="tool"}) has appeared.
   
5. Every assistant message must obey these rules.

\vspace{1em}
\textbf{Available tools}

\texttt{MCP\_TOOL\_LIST}

\vspace{1em}
Question: 

\texttt{QUESTION}
\end{promptbox}

\end{document}